\documentclass[a4paper]{aa}
\usepackage[latin1]{inputenc}
\usepackage{amsmath}
\usepackage{amsfonts}
\usepackage{amssymb}
\usepackage{graphicx}
\usepackage{txfonts}  
\usepackage{natbib}
\begin{document}

\bibliographystyle{aa}

\title{A weak-lensing analysis of the Abell 2163 cluster
\thanks{Based on observations obtained with MegaPrime/MegaCam, a joint project 
of CFHT and CEA/DAPNIA, at the Canada-France-Hawaii Telescope (CFHT) which is 
operated by the National Research Council (NRC) of Canada, the Institute 
National des Sciences de l'Univers of the Centre National de la Recherche 
Scientifique of France, and the University of Hawaii.}
}
\author{M. Radovich \inst{1}  \and E. Puddu \inst{1}  \and A. Romano \inst{1}
\and A. Grado \inst{1} \and F. Getman\inst{2}}
\offprints{M. Radovich: radovich@oacn.inaf.it}

\institute{INAF - Osservatorio Astronomico di Capodimonte, via Moiariello 16, I-80131, Napoli 
\and INAF - VSTceN, via Moiariello 16, I-80131, Napoli}

\date{received; accepted}

 \abstract
   {}
   {We attempt to  measure the main physical properties (mass, velocity dispersion, and total luminosity) 
   of the cluster Abell 2163. }
   {A weak-lensing analysis is applied to a deep, one-square-degree, $r$-band CFHT-Megacam image of the Abell 2163 field. The observed shear is fitted with Single Isothermal Sphere and Navarro-Frenk-White models to obtain the velocity dispersion and the mass, respectively; in addition, aperture densitometry is used to provide a mass estimate at different distances from the cluster centre. The luminosity function is  derived, which enables us to estimate the mass/luminosity ratio.
 }
   {Weak-lensing analyses of this cluster, on smaller scales, have produced results that conflict with each other. The mass and velocity dispersion obtained in the present paper are compared and found to agree well with  values computed by other authors from X-ray and spectroscopic data. }
   {}


\maketitle

\section{Introduction}
\label{sec:intro} 

\begin{figure}
\includegraphics[width=9cm]{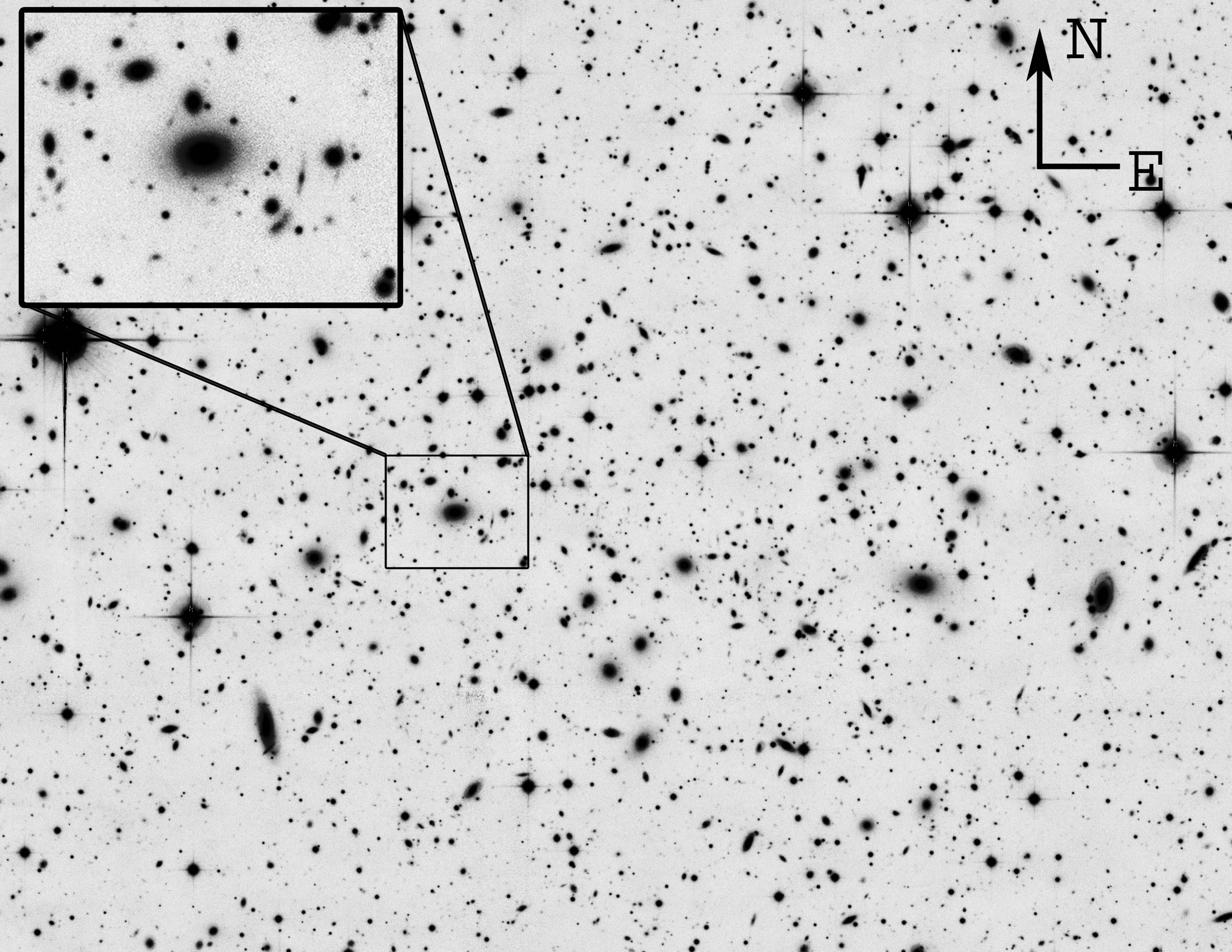} 
\caption{$r-$band image of the Abell 2163 field. The zoomed image shows the bright elliptical galaxy identified as the centre of the cluster from  X-ray maps: also visible are the faint gravitational arcs.\label{fig:image}}
\end{figure}

\object{Abell 2163} is a  cluster of galaxies at $z$=0.203  of richness class
2 \citep{ApJS...70....1A} and without any central cD galaxy (Fig.~\ref{fig:image}). 
It is  one of  the hottest clusters known so far with  an X-ray temperature of 14 keV and
an  X-ray luminosity  of $6 \times 10^{45}$  erg  s$^{-1}$, based  on Ginga  satellite
measurements \citep{apj...390..345a}.  \citet{aa...293..337e} used ROSAT/PSPC  and GINGA
data to map the gas distribution; they showed that the gas  extends to at least 4.6 Mpc or
15 core radii and is elongated in the east-west direction; 
they estimated a  total mass  $(1.43\pm0.05) \times 10^{15} M_{\odot}$ $(h=0.5)$
 inside that radius, which is 2.6 times higher than the total mass  of Coma. The
corresponding gas mass  fraction, 0.1 $h^{-3/2}$, is typical  of  rich
clusters.  The peak of the X-ray emission was found to be close to 
a bright elliptical galaxy ($\alpha=16^h 15^m 49.0^s$, 
 $\delta=-06^\circ 08\arcmin 41\arcsec $), which was confirmed by later X-ray observations \citep{apj...664..761m}. 
Two faint gravitational arcs are visible close to this galaxy (Fig.~\ref{fig:image}); 
 the redshift of the source galaxies is $z_s \sim 0.73$ \citep{apj...449...18m}.
The gas velocity dispersion is also very high, $\sigma = 1680$  km s$^{-1}$ \citep{arnaud94};
\citet{apj...664..761m} derived a velocity dispersion of $\sigma = 1381 \pm 324$ km/s from spectroscopic data.

ASCA observations of  Abell 2163  \citep{apj...456..437m} measured a dramatic  drop in 
the temperature  at large radii: this placed strong constraints on  the total mass
profile,  assumed  to follow  a  simple  parametric  law \citep{apj...456..437m}.
Considerable gas temperature variations  in the central 3-4 core radii
region were also found. The total mass  derived inside $0.5 h^{-1}$ Mpc was $(4.3 \pm 0.5) \times 10^{14} h^{-1}$ $M_{\odot}$, 
while inside $1.5 h^{-1}$ Mpc it was found to be $(1.07 \pm 0.13) \times 10^{15} h^{-1}$ $M_{\odot}$. 

Abell 2163   is remarkable also in the radio band: as first reported by   \citet{aas...185.5307H},  
it shows  a very extended and powerful radio halo. \citet{aa...373..106f} further investigated
the radio properties of the cluster. In addition to its size ($\sim 2.9$ Mpc), the halo is slightly
elongated in the E-W direction; the same elongation is also seen in the X-ray. 

All of this evidence indicates that  the  cluster  is  unrelaxed and   has experienced 
a recent or is part of an ongoing merger of two large clusters 
\citep{aa...293..337e,  aa...423..111f}.  
This was confirmed by \citet{maurogordato08}, who interpreted the
properties of Abell 2163 in terms of a recent merger, in which the main component is positioned in the EW direction and a further
northern subcluster (Abell 2163-B) is related to the same complex. They  used optical and spectroscopic data
to compute, in addition, the virial mass, $M_{\rm vir} = (3.8 \pm 0.4) \times 10^{15}$ $M_\odot$ and the gas velocity dispersion, $\sim 1400$ km s$^{-1}$.

\citet{apj...482..648s} first performed a weak-lensing analysis of Abell 2163 using a $2048\times2048$
CCD at the prime focus of the Canada-France Hawaii Telescope (CFHT). They mapped the dark matter distribution
up to $7\arcmin$ ($\sim 1 h^{-1}$ Mpc); the mass map showed two peaks, one close to the elliptical galaxy, the other at $3\arcmin$ W. 
The mass obtained by weak lensing alone was a 
 factor of $\sim 2$ lower than that derived from X-ray data: they interpreted the  discrepancy in mass measurement as the result of an extension of the mass distribution, beyond the edges of the CCD frame; taking this effect into account, a reasonable agreement is achieved between the mass determined by X-ray and weak lensing. 
A fit of the shear profile with that expected for a singular
isothermal model provided a velocity dispersion measurement of $\sigma = 740$ km s$^{-1}$, which was lower than the expected value $\sigma >1000$ km s$^{-1}$.

\citet{apj...613...95c} completed a weak-lensing analysis of Abell 2163 using FORS1 at the VLT in imaging mode. They measured a higher velocity dispersion of $\sigma =  1021 \pm 146$ km s$^{-1}$. They explained their disagreement with \citet{apj...482..648s}
 by the fact that those authors used a bright cut ($V > 22$ mag, $I> 20.5$ mag) for the selection of background galaxies, whereas they chose $R > 23.3$ mag. 

Wide-field cameras, such as the ESO Wide-Field Imager (WFI) with a field of view of $34\arcmin \times 33\arcmin$, and
the Megacam camera mounted at the CFHT ($\sim 1$ square degree), are particularly well suited for the weak-lensing study of clusters because they enable the clusters to be imaged well beyond their radial extent.
We use public archive data of Abell 2163, acquired using the Megacam camera, to complete a revised weak-lensing analysis of this cluster and derive the luminosity function of the cluster galaxies.

This paper is organized as follows. Section~\ref{sec:data} describes the data and  steps followed in the reduction. The weak-lensing analysis and  
determination of  mass are discussed in Sect.~\ref{sec:analysis}.  Finally, the cluster luminosity function is derived and the mass to luminosity ratio is computed in Sect.~\ref{sec:lumfunc}.

We adopt $H_0 = 70$ km s$^{-1}$ Mpc$^{-1}$, $\Omega_\Lambda = 0.7$, $\Omega_m = 0.3$, which corresponds to a linear scale of 3.34 kpc/$\arcsec$ at the redshift of Abell 2163.

\section{Observations and data reduction}
\label{sec:data}

\begin{figure}
 \centering
 \includegraphics[width=8 cm]{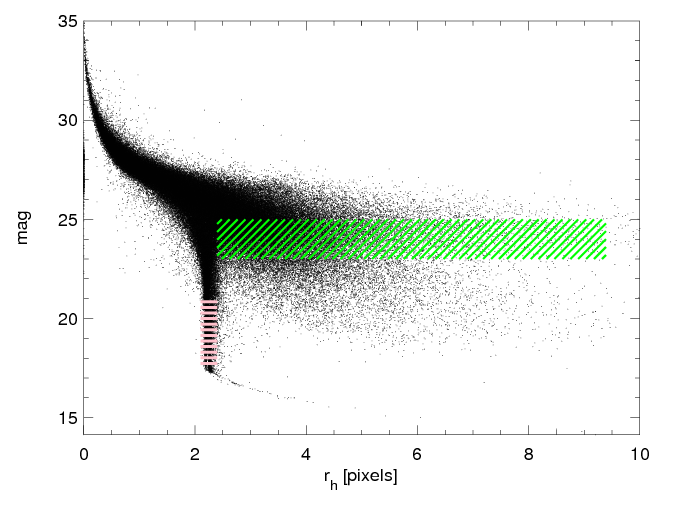}
 \caption{Selection of galaxies used for the lensing analysis (area with oblique lines) and of the stars used for the 
PSF correction (area with horizontal lines) in the  CFHT-Megacam $r$-band image; $r_h$ is the half-light radius. }
 \label{fig:pcats}
\end{figure}

Abell 2163 was observed in 2005 with the Megacam camera at the 3.6m Canada-France Hawaii Telescope in the $r$-band,  with a total exposure time of 2.7hr. The prereduced (bias and flat-field corrected) images were retrieved from the Canadian Astronomy Data Centre 
archive\footnote{http://www4.cadc-ccda.hia-iha.nrc-cnrc.gc.ca/cadc/}.
Before coadding the different exposures, it was necessary to remove the effect of distortions produced by the optics and by the telescope. This was completed   using the \textsc{AstromC} package, which is a porting to C++ of the \textsc{Astrometrix} package described in \citet{VIRMOSU};  we refer the reader to this paper for further details. For each image, an astrometric solution was computed, assuming  the USNO-A2 catalog as the astrometric reference and  taking into account the positions of the same sources in the other exposures. The absolute accuracy of the astrometric solutions with respect to the USNO-A2 is limited to its nominal accuracy, $\sim 0.3$\arcsec; the internal accuracy of the same sources detected in different images is far lower ($\sim 0.01$\arcsec), which enables us to optimize the Point Spread Function (PSF) of the final coadded image.   \textsc{AstromC} for each exposure computed  an offset to the zero point to take into account changes e.g. in the  transparency of the atmosphere, relative to one exposure that was taken as reference. Photometric zero points were given already in the header of the images; Table~\ref{tab:photcal} summarizes the photometric parameters. All  images were resampled according to the astrometric solution and coadded together using the \textsc{SWarp} software developed by E. Bertin\footnote{http://terapix.iap.fr/}. Finally, catalogs of sources were extracted using \textsc{SExtractor}.  Galaxies and stars were selected
by the analysis of the $r_h$ versus magnitude diagram, where $r_h$ is the half-light radius (Fig.~\ref{fig:pcats}).
The coadded image was inspected to search for regions with spikes and halos around bright stars. Such regions
were masked on the image with  \textsc{DS9}\footnote{http://hea-www.harvard.edu/RD/ds9/} and sources inside them were discarded from the catalog. In addition, we did not use the outer part of the image, where the PSF rapidly degraded. The residual available area is 1775 ${\rm arcmin}^2$.

\begin{table*}
\caption{Photometric calibration terms and magnitude limits computed for point-like sources at different signal to noise ratios  in the CFHT-Megacam $r$-band. $A_r$ is the average Galactic extinction, for $E(B-V)=0.35$.
Megacam zero points are defined so that  magnitudes are already on the AB system, and are given here such that the airmass is 0. 
\label{tab:photcal}}
\centering
\begin{tabular}{cccccc c c c c c c}
\hline\hline
Zero point & Color term & Extinction coeff.  & $A_r$ & $m_{\rm AB}(\sigma=3)$ & $m_{\rm AB}(\sigma=5)$ & $m_{\rm AB}(\sigma=10)$\\
\hline 
26.1 & 0.00$\times$ (g-r) & 0.10 &   0.92  &  27.0 & 26.4 & 25.5\\ 
\hline 
\end{tabular} 
\end{table*}

We note that Abell 2163 is located in a region of high Galactic extinction: from the maps by \citet{Schlegel}, using the \textsc{dust\_getval}  code\footnote{http://www.astro.princeton.edu/$\sim$schlegel/dust/} 
we obtain $0.27 < \rm{E(B-V)} < 0.43$ in the field, with an average value of  $\rm{E(B-V)}=0.35$. 
Such change is significantly higher than the typical uncertainty  in $\rm{E(B-V)}$ \citep[$\sim$ 16\%,][]{Schlegel}: it is therefore more appropriate to correct the magnitude of each galaxy for the extinction value at its position, rather than using the same average value.

\begin{figure}
 \centering
 \includegraphics[width=10 cm]{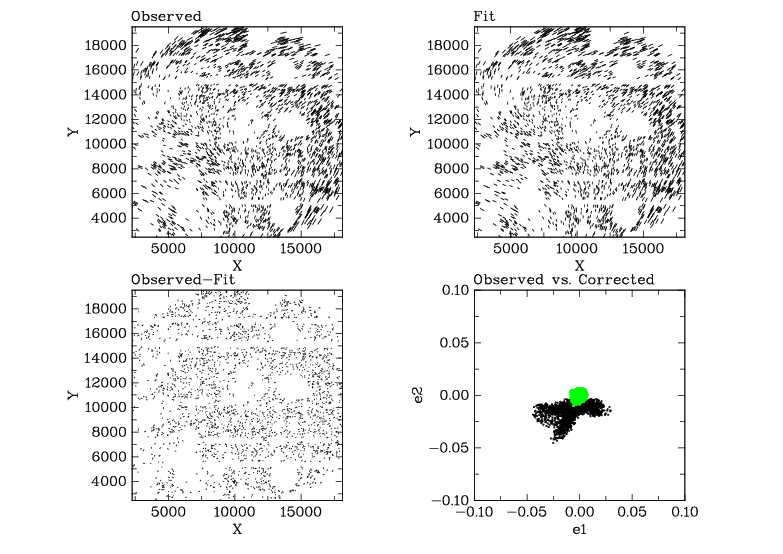}
 \caption{PSF correction: the first three panels show the spatial pattern of the observed, fitted and residual ellipticities of stars: a scaling factor was applied for display purposes; $X$ and $Y$ are the pixel coordinates in the image. 
The last panel shows the observed versus corrected ellipticities.}
 \label{fig:psf}
\end{figure}

\begin{figure}
 \centering
 \includegraphics[width=9 cm]{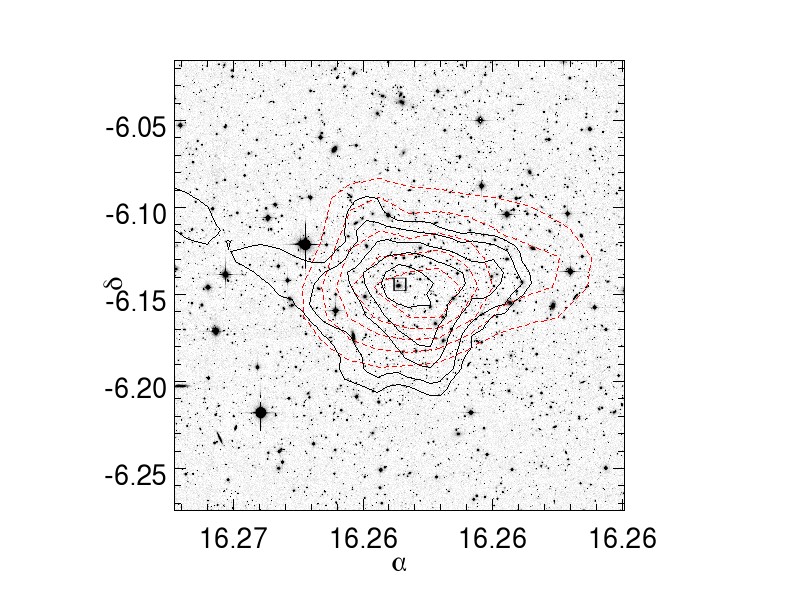}
  \includegraphics[width=9 cm]{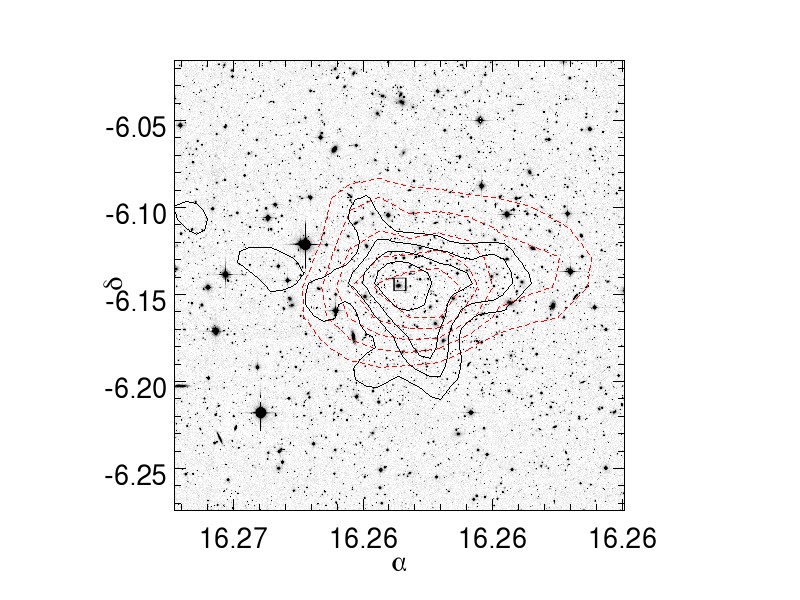}
 \caption{\textit{S-maps} obtained by aperture densitometry and convolution with {\em up:} a Gaussian 
filter function (size: 5 arcmin); {\em down:} the filter function proposed by \citet{schirmer_phd}. The contour levels are plotted at $S=(3,4,5,6,7)$, where $S$ is defined in Sec.~\ref{subsubsec:massap}. The small box 
indicates the position of the elliptical galaxy with arcs. The dashed contours show for comparison the density distribution of cluster galaxies \citep[see also][]{maurogordato08}.}
 \label{fig:apmass}
\end{figure}

\begin{figure}
 \centering
 \includegraphics[angle=90, width=8 cm]{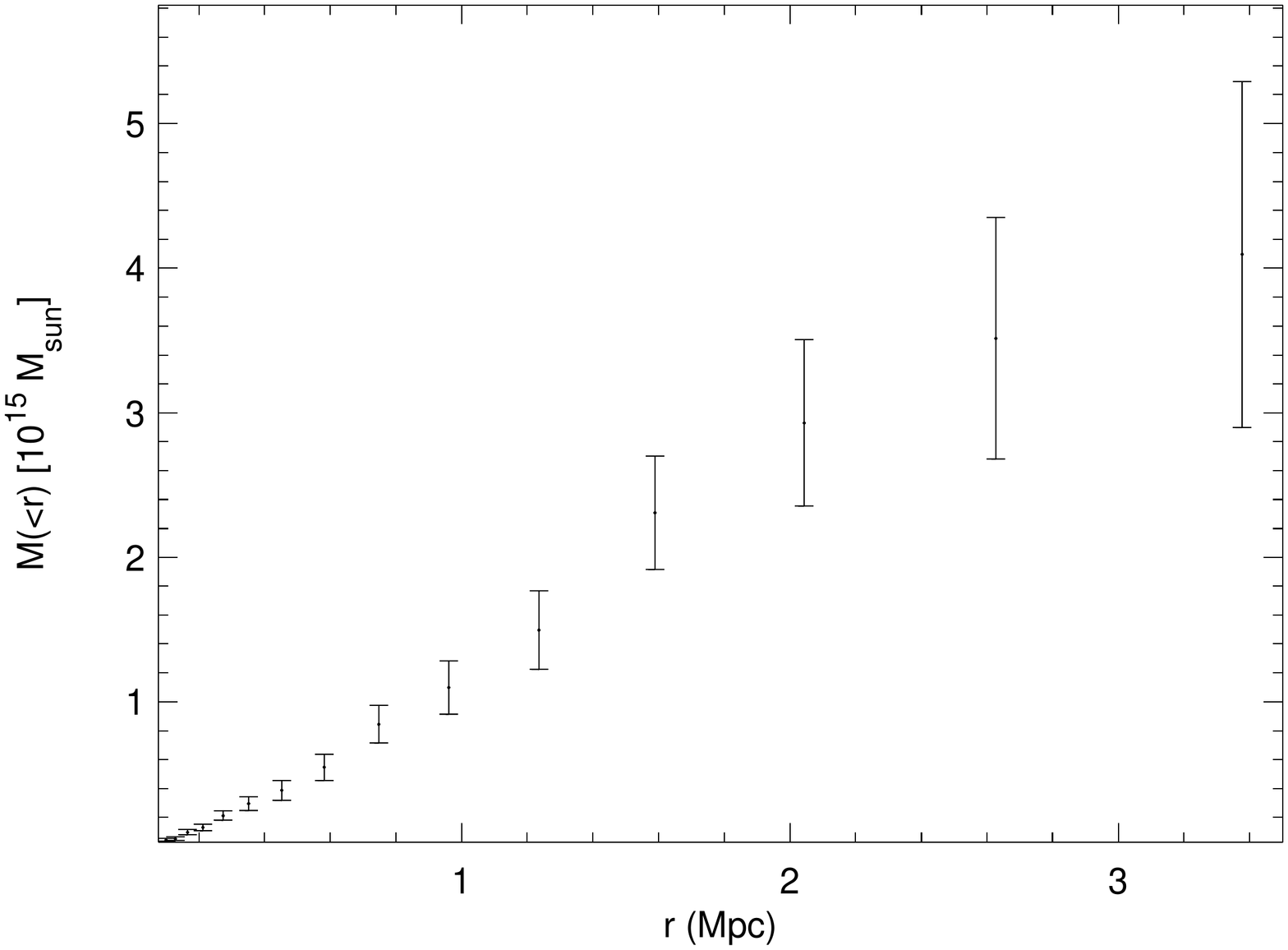}
 \caption{Mass profile obtained by aperture densitometry.}
 \label{fig:apdens}
\end{figure}

\section{Weak-lensing analysis}
\label{sec:analysis}

Weak-lensing relies on the accurate measurement of the average distortion produced by a distribution of
matter  on the shape of background galaxies. As the distortion is small, the removal of systematic effects, in particular the effect of the PSF both from  the telescope and from the atmosphere,  is of crucial importance. Most of 
published weak-lensing results have adopted the so--called KSB approach  proposed by  \citet{kaiser95} and  \citet{kaiser97}.   We summarize  the main points here and refer to e.g. \citet{kaiser95},  \citet{kaiser97}, and \citet{apj...504..636h} for  more detailed discussions. 

In the KSB approach, for each source the following quantities are computed from the moments of the intensity distribution:
the observed ellipticity $e$, the smear polarizability $P^{\rm sm}$, and the shear polarizability $P^{\rm sh}$.
It is assumed that the PSF can be described as the sum of an isotropic component  (simulating the effect of seeing) and an anisotropic part.
 The intrinsic ellipticity $e_s$ of a galaxy is related to its observed one, $e_{\rm obs}$, and to the shear, $\gamma$, by:

\begin{equation}
 e_{\rm obs} = e_{s} +P^{\gamma} \gamma+P^{\rm sm} p. \label{eq:ellipticity} 
\end{equation} 

The term $p$ describes the  effect of the  PSF anisotropy (starred terms indicate that they are derived from measurement of stars):
\begin{equation}
p = e^*_{\rm obs}/P^{\rm sm*}.
\end{equation}
It is necessary to fit this quantity as it  changes with the position in the image,  using e.g. a polynomial such that it can be
extrapolated to the position of the galaxy. In our case, we find that a polynomial of order 2 describes the data well.

The term $P^{\gamma}$, introduced by \citet{kaiser97} as the {\em pre--seeing shear polarizability}, describes the effect of  seeing
and is defined to be:
\begin{equation}
 P^{\gamma} = P^{\rm sh} - P^{\rm sm} \frac{P^{\rm sh*}}{P^{\rm sm*}} .
\end{equation}
As discussed by  \citet{apj...504..636h}, the quantity $ \frac{P^{\rm sh*}}{P^{\rm sm*}}$ should be computed with the same weight function used for the galaxy to be corrected. For this reason, the first step is to compute its value using weight functions of size drawn from a sequence of bins in the half-light radius $r_h$.  
In many cases,  $ \frac{P^{\rm sh*}}{P^{\rm sm*}}$ can be assumed to be constant across  the image and
be computed from the average of the values derived from the stars in the field. 
We preferred to fit the quantity for the Megacam image, considering its size, as a function in addition of the coordinates (x, y), using a polynomial of order 2.
For each galaxy of size $r_h$,
we then assumed the coefficients computed in the closest bin to finally derive the value of $\frac{P^{\rm sh*}}{P^{\rm sm*}}$. 

The implementation  of the KSB procedure is completed using a modified version of  Nick Kaiser's IMCAT tools, kindly provided to us by T. Erben \citep[see][]{aa...468..859h}; these tools enable measurement of  the quantities relevant to the lensing analysis, starting from catalogs obtained using \textsc{SExtractor}. The package also enables us to separate stars and galaxies in the $r_h$-mag space and compute the PSF correction coefficients $P^\gamma$, $p$. In addition, we introduced the possibility to fit $P^\gamma$ versus the coordinates (x,y), as explained above,  and for each galaxy used the values of both 
$P^\gamma$ and $p$ computed in the closest bin of $r_h$.

Stars were selected in the range $17.5 {\rm mag} < r < 21 {\rm mag}$, $ 0.37\arcsec < r_h < 0.45 \arcsec$, providing 2400 stars usable to derive the quantities needed for the PSF correction.  
As discussed above, these quantities were fitted with a polynomial both for PSF anisotropy  and seeing correction: we verified that the behaviour of the PSF  across the CCDs enabled  
a single  polinomial function to be used for the entire image (Fig.~\ref{fig:psf}).  
Galaxies used for shear measurement were selected using the following criteria:  
$P_\gamma > 0.25$, $\nu_{\rm max} > 5$ , $r_h > 0.45\arcsec$, $23 \rm{mag} < r < 25 \rm{mag}$, and ellipticities smaller than one. 
We finally obtained approximately 17000 galaxies, which implied that the average density of galaxies in the catalog was $\sim 8$ galaxies/arcmin$^2$.

The uncertainty in ellipticities was computed as in \citet{apj...532...88h}:
\begin{equation}
w=\frac{1}{\sigma_\gamma^2}=\frac{P^{\gamma^2}}{P^{\gamma^2}  \sigma_{e_0}^2  +
 \left\langle  \Delta e ^2   \right\rangle }, \label{eq:well}
\end{equation}
where $\left\langle  \Delta e^2   \right\rangle^{1/2}$ was the uncertainty in the measured ellipticity, $\sigma_{e_0} \sim 0.3$ was the typical intrinsic rms of galaxy ellipticities.

\begin{table*}
\caption{Best-fit values obtained from the fit of the shear from the NFW model ($c = 4.09$). For comparison, the lower rows give the masses computed at the same radii from the best-fit SIS model ($\sigma_v = 1139_{-56}^{+53}$ km/s,  $\theta_e =22 \arcsec$)
and from aperture mass densitometry (A.D.).\label{tab:masses}}
\centering
\begin{tabular}{l cc cc cc cc}
\hline\hline
&
$r_{\rm vir}$ & $M_{\rm vir}$ &
$r_{\rm 200}$ & $M_{\rm 200}$ &
$r_{\rm 500}$ & $M_{\rm 500}$ &
$r_{\rm 2500}$ & $M_{\rm 2500}$ \\

&
(kpc) & $(10^{14}  M_{\odot})$ &
(kpc) & $(10^{14}  M_{\odot})$ &
(kpc) & $(10^{14}  M_{\odot})$ &
(kpc) & $(10^{14}  M_{\odot})$ \\

\hline
NFW &
$3012\pm170$ &  $22\pm4 $ & 
$2362\pm130$  &  $18\pm3 $ & 
$1514\pm90$  &  $12\pm2$   &  
$ 623\pm40$  & $4.2\pm0.7$   \\

SIS &
 & $18\pm5$ &  & $14\pm4$ & & $9\pm2$ & & $4\pm1$ \\ 

A.D. &
 & $  38 \pm 10 $ &
 & $  33 \pm 7 $ &
 & $  20 \pm 3$ &
 & $  5.8 \pm 0.8$ \\

\hline 
\end{tabular} 

\end{table*}

\subsection{Mass derivation}
\label{subsec:massder}
Weak lensing measures the {\em reduced shear} $g=\tfrac{\gamma}{(1-\kappa)}$. The convergence $\kappa$ is defined by $\kappa=\Sigma/\Sigma_{\rm crit}$, where $\Sigma$ is the surface mass density and $\Sigma_{\rm crit}$ is the critical
surface density:
\begin{equation}
\Sigma_{\rm crit} = \frac{c^2}{4\pi G} \frac{D_s}{D_l D_{ls}} = \frac{c^2}{4\pi G} \frac{1}{D_l \beta} ,
\end{equation}
 $D_{ls}$, $D_s$, and $D_l$ being the angular distances between lens and source, observer and source, and observer 
and lens respectively. In the weak lensing approximation, $\kappa \ll 1$, so that $g \sim \gamma$. However,
the measured value of $\kappa$ includes an unknown additive constant (the so-called {\em mass-sheet degeneracy}): 
this degeneracy can be solved by 
assuming either that $\kappa$ vanishes at the borders of the image, or a particular mass profile for which the expected
shear is known. Both approaches are used here.

For Abell 2163, $\Sigma_{\rm crit} = 9.69 \times 10^{13} \beta^{-1} M_{\odot}$ {\rm arcmin}$^{-2}$.
In our case, we were unable to assign a redshift to the source galaxies;  we however  assumed the {\em single-sheet approximation}, which implies  that the background galaxies lie at the same redshift \citep{aa...369....1k}.  To compute the value of the redshift, we used the publicly-available photometric
redshifts obtained by \citet{aa...457..841i} for the VVDS F02 field with Megacam photometric data. We applied the
same cuts adopted here for the $r$-band magnitude data and assumed a Gamma probability distribution \citep{aa...422..407g}:
\begin{equation}
 n(z) = z^{a-1} \frac{\exp(-z/z_s)}{\Gamma(a) z_s^a},
\end{equation}
We found that $a=2.04$ and $z_s=0.57$ were the best-fit parameters, and $\rm{Med}(z) = 0.98$. We therefore adopted a median
redshift of $z \sim 1$, which provided  $\left\langle \beta  \right\rangle = 0.58$.

\begin{figure}
 \centering
 \includegraphics[width=9cm]{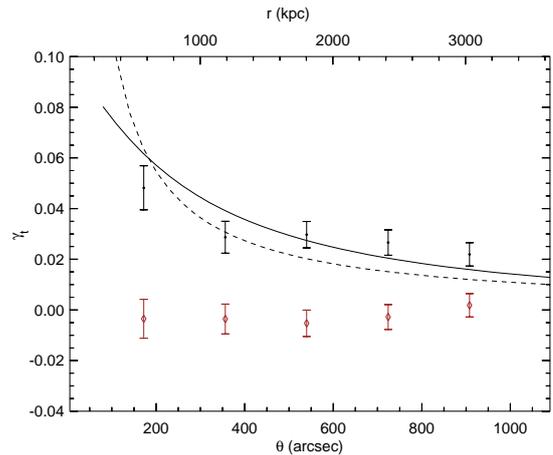}
 \caption{The tangential component of the shear (dark points) is displayed as a function of the distance from the assumed centre of the cluster. The lines show the result of the best-fit to the unbinned data using: a NFW profile with $c_{\rm vir}$ given by Eq.~\ref{eq:cvirmass} (solid line), and a SIS profile (dashed line).  Also shown is the radial component of the shear (diamonds), which is expected to be null in the absence of systematic errors.}
 \label{fig:mnfw}
\end{figure}

\subsubsection{Mass aperture maps}
\label{subsubsec:massap} 

Figure~\ref{fig:apmass} displays the \textit{S-map} introduced by \citet{aa...420...75s}, that is:
\begin{eqnarray}
M_{\rm ap} = \frac{\sum_i  e_{t,i} w_i Q(\vert \theta_i-\theta_0 \vert )}{\sum_i{w_i}}\\
\sigma^2_{M_{\rm ap}} = \frac{\sum_i e^2_{t,i} w^2_i Q^2(\vert \theta_i-\theta_0 \vert)}{2(\sum_i{w_i})^2},
\end{eqnarray}
where $e_{t,i}$ are  tangential components of the lensed-galaxy ellipticities computed by considering to be the centre the position
in the grid, $w_i$ to be the weights defined in Eq.~\ref{eq:well}, and $Q$ the filter function discussed below.
The ratio $S=M_{\rm ap}/\sigma_{M_{\rm ap}}$, defined as the  {\em S-statistics} by \citet{aa...420...75s}, provides a direct
estimate of the signal-to-noise ratio of the halo detection. 

For the window function, we tested two possible forms: a Gaussian function and a function close to a NFW profile. 

The Gaussian window function is defined by:
\begin{equation}
Q(\vert \theta-\theta_0 \vert) = \frac{1}{\pi \theta^2_c} \exp \left( -\frac{(\theta-\theta_0)^2}{\theta^2_c}\right),
\end{equation}
where $\theta_0$ and $\theta_c$ are the centre and size of the aperture.

\citet{schirmer_phd} proposed a filter function, whose behaviour is close to that expected from a NFW profile:

\begin{equation}
Q(x)=\left( 1+e^{a-bx} +e^{-c+dx} \right)^{-1} \frac{\tanh (x/x_c)}{\pi \theta_c^2 (x/x_c)},
\end{equation}
where $x=(\theta-\theta_0)/\theta_c$, and we adopted the following parameters: $a=6$, $b=150$, $c=47$, $d=50$, $x_c=0.15$ \citep{aa...442...43h}.

In both cases, we found consistently that (i) the peak of the lensing signal was coincident with the position of the bright elliptical galaxy with  arcs (BCG hereafter), confirming that this was in fact the centre of the mass distribution as  indicated by the X-ray maps; and
(ii) the weak-lensing signal is elongated in the E-W direction. A comparison with Fig.~5 and Fig.~9 in \citet{maurogordato08} indicates that
the mass distribution follows the density distribution of early-type cluster galaxies (the A1 and A2 substructures).

\subsubsection{Aperture densitometry}
\label{subsubsec:apdens} 

We first  estimate the mass profile of the cluster by computing the $\zeta$ statistics \citep{fahlman, apj...497l..61c}:
\begin{eqnarray}
 \zeta (\theta_1) =
  \bar{\kappa} (\theta \le \theta_1) - \bar{\kappa} (\theta_2 < \theta \le \theta_{\rm max}) =
2  \int^{\theta_2}_{\theta_1} \left\langle \gamma_T \right\rangle d \ln \theta \\ \nonumber 
+  \frac{2}{1 - (\theta_2 /\theta_{\rm max})^2} \int^{\theta_{\rm max}}_{\theta_2} \left\langle \gamma_T \right\rangle d \ln \theta \label{eq:apmass} 
\end{eqnarray}

The quantity $M_{\rm ap}(\theta_1) = \pi \theta_1^2 \zeta (\theta_1) \Sigma_{\rm crit}$ provides a lower limit
to the mass inside the radius $\theta_1$, unless $\bar{\kappa} (\theta_2, \theta_{\rm max}) = 0$. 
This formulation of the $\zeta$ statistics is particularly convenient 
because it enables a choice of control-annulus size ($\theta_2$, $\theta_{\rm max}$) that satisfies this condition reasonably well; in addition, the mass computed inside a given aperture is independent of the mass profile of the cluster \citep{apj...497l..61c}.
\citet{clowe04} discussed how the mass estimated by aperture densitometry is affected by asphericity and projected substructures in clusters, as in the case of Abell 2163: they found that the error was less than 5\%.

The cluster X-ray emission was detected out to a clustercentric radius of $2.2 h^{-1}$ Mpc \citep{apj...482..648s}, which corresponds to $\sim 900\arcsec$.
We took advantage of the large available area and chose  $\theta_2 =1300\arcsec $, 
$\theta_{\rm max} \sim 1500\arcsec$, which provided $\sim$ 3000 sources in the control annulus. 
The mass profile is displayed in Fig.~\ref{fig:apdens}; the mass values computed at different radii are shown in Table~\ref{tab:masses}.

\subsubsection{Parametric models}
\label{subsubsec:models} 

We consider a Singular Isothermal Sphere (SIS) and a Navarro-Frenk-White (NFW) mass profile,
for which the expected shear can be expressed analytically. The fitting of the models is completed by minimizing the  
log-likelihood function \citep{aa...353...41s}:

\begin{equation}
 l_\gamma = \sum_{1=1}^{N_\gamma}{\left[ \frac{\vert\epsilon_i-g(\theta_i)\vert^2}{\sigma^2[g(\theta_i)]} 
+ 2 \ln \sigma[g(\theta_i)]\right] }, \label{eq:ll}
\end{equation}
with $\sigma[g(\theta_i)] = (1-g(\theta_i)^2) \sigma_e$.

In the case of a SIS profile, the shear is related to the velocity dispersion $\sigma$ by:
\begin{equation}
 \gamma_t(\theta) = \frac{2\pi}{\theta} \frac{\sigma^2}{c^2} \frac{D_{ls}}{D_s} = 
\frac{\theta_E}{\theta} 
\end{equation}

For the Navarro-Frenk-White (NFW) model, the mass profile is \citep{apj...534...34w}: 
\begin{equation}
\rho(r) = \frac{\delta_{c} \rho_c}{(r/r_{s}) (1+r/r_{s})^2},
\end{equation}
where
$\rho_c=3 H^2(z)/(8 \pi G)$ is the critical density of the universe at the cluster redshift;
$r_s$ is a characteristic radius related to the virial radius by the concentration parameter 
$c_{\rm vir}=r_{\rm vir}/r_s$; $\delta_c$ is a characteristic overdensity of the halo:
\begin{eqnarray}
\delta_c = \frac{\Delta_{\rm vir}}{3} \frac{c^3}{ln(1+c)-c/(1+c)},\\ \nonumber
\Delta_{\rm vir} \sim (18\pi^2 + 82 (\Omega_M(z)-1) - 39(\Omega_M(z)-1)^2)/\Omega_M(z). 
\end{eqnarray}
The mass of the halo is:
\begin{equation}
 M_{\rm vir}=\frac{4}{3}\pi \Delta_{\rm vir} \rho_c r^3_{\rm vir}
\end{equation}

 \citet{mnras.321..559b} used simulations of clusters to show that  the virial mass and the concentration are linked by the relation:

\begin{equation}
c_{\rm vir} =\frac{K}{1+z} \left(\frac{M_{\rm vir}}{M_\star}   \right)^\alpha,
\label{eq:cvirmass} 
\end{equation}
with $M_\star= 1.5 \times 10^{13} h^{-1}$ $M_{\odot}$, $K = 9$, $\alpha=-0.13$.

\citet{mnras.379..190c} computed the values of $K$ and $\alpha$  fitting Eq.~\ref{eq:cvirmass} to the values of virial mass and 
concentration measured in a sample of 100 clusters; they adopted $M_\star = 1.3 \times 10^{13} h^{-1}$ $M_{\odot}$ and found $K=14.5 \pm 6.4$,  $\alpha = -0.15 \pm 0.13$, which provides values of the concentration which are approximately 1.6 higher than obtained using the relation proposed by  \citet{mnras.321..559b}.  Given the large uncertainty in the value of $K$, we preferred to adopt the values of \citet{mnras.321..559b}. 
We used the expression of the  shear $\gamma_{t}(r)$ derived by \citet{aa...313..697b}; the minimization in Eq.~\ref{eq:ll} was completed using 
the MINUIT package. Figure~\ref{fig:mnfw} shows the results of the fit and, for comparison, the binned values of the tangential and radial components of the shear: these are consistent with zero, as expected in the absence of systematic effects. 
 Table~\ref{tab:masses} shows the masses obtained by model fitting (SIS and NFW), as well as those
obtained by aperture mass densitometry at different distances from the BCG, which was  assumed to be the centre of the cluster.
In addition to $M_{\rm vir}$, the masses obtained for $\rho/\rho_c = 200, 500, 2500$ and the corresponding  radii $r_{200}$, $r_{500}$ and $r_{2500}$, are also displayed.  The value of the virial mass, $M_{\rm vir} =  (22\pm 4) \times 10^{14} M_{\odot}$, confirms Abell 2163 as a massive cluster compared to other clusters \citep{mnras.379..190c}.

\begin{figure}
 \centering
 \includegraphics[viewport=20 150 565 705,width=9cm]{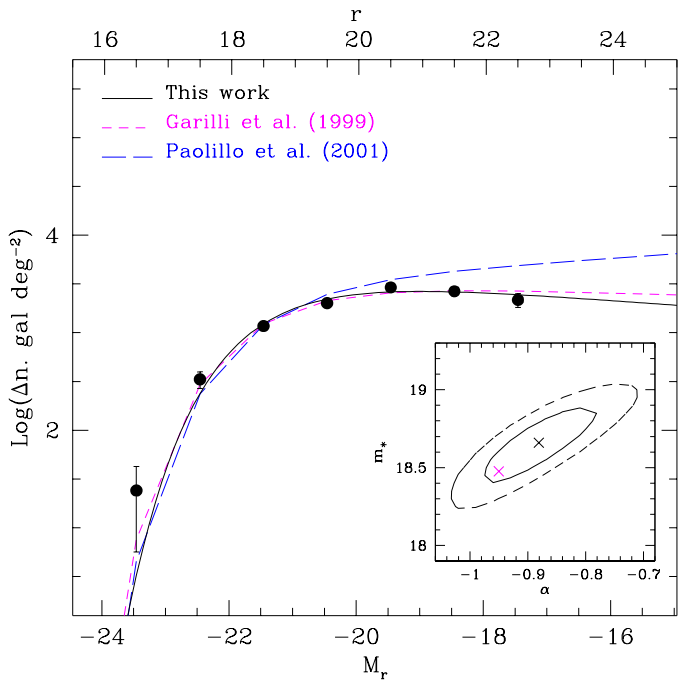}
 \caption{The $r$-band LF of Abell 2163 compared with other determinations in the  literature. }
 \label{fig:lf}
\end{figure}

\begin{figure}
 \centering
 \includegraphics[viewport=15 150 575 700,width=9cm]{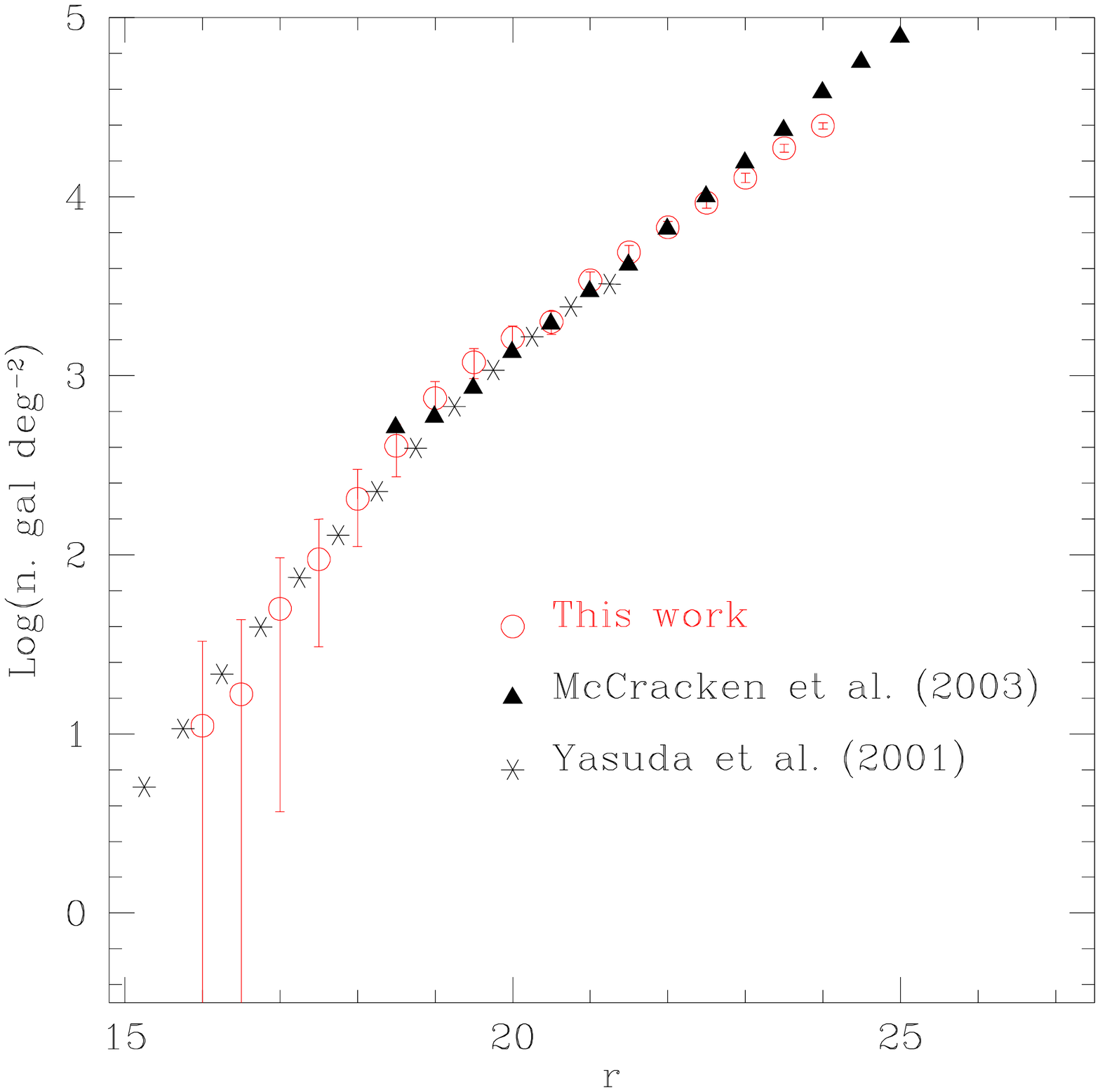}
 \caption{
Galaxy counts derived from the control field (empty red circles), compared to literature. 
The triangles represent the deep counts from the CFH12K-VIRMOS field \citep{AA.410.17M} 
corrected to the $r_{AB}$ according to \citet{PASP.107.945F}; 
the stars mark the counts from the SDSS commissioning data \citep{AJ.122.1104Y}. 
The error bars of this work take into account only the Poissonian 
errors ($\sqrt{n}$/area), whereas for the literature counts the errors are not displayed because they are smaller than the point dimensions. 
}
 \label{fig:counts}
\end{figure}

\begin{figure}
 \centering
 \includegraphics[viewport=15 150 575 435,width=9cm]{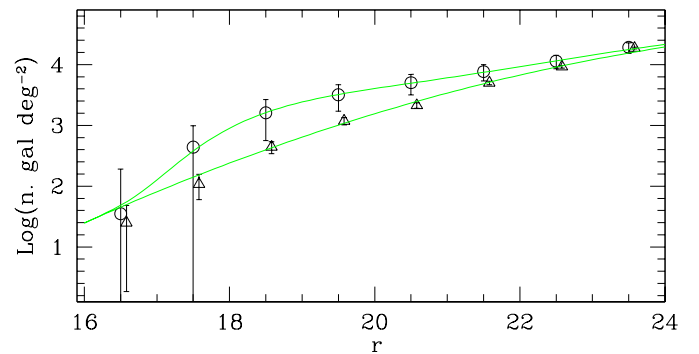}
 \caption{Galaxy counts in the Abell 2163 line of sight (circles) and in the control field 
(triangles: for display purposes, an offset in $r$ was applied to these points). 
The lines show the result of the joint fit. The errors bars were computed as described in the text.
}
 \label{fig:dcounts}
\end{figure}

\begin{table}
\caption{Best-fit parameters and errors. $a$, $b$, and $c$ describe the shape of 
galaxy counts, whereas $\alpha$, $m_{\star}$, and $\Phi_{\star}$ describe the shape of 
the cluster LF. $a$, $b$, $c$, and $\Phi_{\star}$ are in units of $deg^{-2}$. 
The last two columns show the half width of the 68\% and 95\%  
confidence range, with two degrees of freedom. \label{tab:lf}}
\centering
\begin{tabular}[c]{llll}
\hline\hline
& Best-fit & 68\% & 95\%\\
\hline
$\alpha$ & -0.88 & 0.09 & 0.15  \\ 
$m_{\star}$    & 18.66 & 0.23 & 0.40  \\ 
$\Phi_{\star}/10^3$ & 3.8 &   &  \\ 
$a$            & 3.19 &  &  \\ 
$b$            & 0.36 &  &    \\ 
$c$            & -0.022 &  &   \\
\hline
\end{tabular}
\end{table} 

\section{Luminosity function}
\label{sec:lumfunc}

To compute the total $r-$band luminosity of the cluster and hence the $M/L$ ratio, we first derived its luminosity function (LF hereafter). The $r-$band magnitudes were corrected
for Galactic reddening as explained in Sect.~\ref{sec:data}; no k-correction was applied because it is negligible 
at the redshift of Abell 2163,  according to \citet{AJ.122.1104Y}.

We defined the cluster region to be the circular area encompassed within $r_{200}$ ($\sim 0.2 \deg$, 
see Table~\ref{tab:masses}) and  centred on the BCG.
Our control field of galaxies was assumed to be those falling into the outer side (0.36 $\deg^2$) of a squared region centered on the BCG of area about 0.25 $\deg^2$.

With this choice, we are confident that we minimize the contamination of cluster galaxies 
and  take into account background non-uniformities in the angular scale of 
the cluster.
Figure~\ref{fig:counts} shows that the $r-$band galaxy counts in the control field are consistent with those found in the literature.

The LF was computed by fitting the galaxy counts in the cluster and control-field areas:
we adopted the rigorous approach introduced by \citet{mnras.360.727a},
which allows us to include, at the same time, in the likelihood function to be minimized, the contribution of both background and cluster galaxies.
As the model for the counts of the cluster field, we used the sum of a power-law (the 
background contribution in the cluster area) and a \citet{Schechter} function, 
normalized to the cluster area $\Omega_{cl}$:

\begin{eqnarray}
p_{cl,i} = \Omega_{cl}  \Phi_{\star}  10^{0.4 (\alpha +1) (m_{\star}-m)}  
\exp(-10^{0.4 (m_{\star}-m)}) \\ \nonumber
+ \Omega_{cl}  10^{a+b (m-20)+c (m-20)^2}.
\end{eqnarray}

For the control field, this reduces to the power-law only, normalized to the background 
area $\Omega_{bkg}$:

\begin{equation}
p_{bkg,i} = \Omega_{bkg} * 10^{a+b (m-20)+c (m-20)^2},
\end{equation}
where $\Phi_{\star}$,$\alpha$, and $m_{\star}$ are the conventional Schechter parameters as usually defined;
$a$, $b$, and $c$ describe the shape of the galaxy counts in the reference-field direction;
and the value of 20 was chosen for numerical convenience.

Best-fitting parameters (see Table~\ref{tab:lf}) were determined simultaneously by using a conventional
routine of minimization on the unbinned distributions. The data were binned for 
display purpose only in Fig.~\ref{fig:dcounts}, which shows the binned galaxy counts
in the control field  (empty triangles) and  cluster (empty circles) areas; the joint fit to the unbinned data sets 
is also overplotted. Error bars are calculated to be $\sqrt{n}/\Omega$.
To check the effect of the uncertainty in the position-dependent extinction correction (see Sect.~\ref{sec:data}), we computed the same parameters in a set of catalogs for which the extinction correction  of each  galaxy was randomly modified within $\pm$ 15\%, the expected uncertainty in $E(B-V)$ \citep{Schlegel}: the rms uncertainty in the parameters derived in this way is negligible compared to the uncertainties in the fitting. 
Figure~\ref{fig:lf} displays the derived LF compared with selected determinations from the literature, which have been  converted to our cosmology.

We compare our LF with \citet{AA.342.408G} and  \citet{AA.367.59P}, which both used data calibrated to the Thuan \& Gunn photometric system;
according to \citet{PASP.107.945F}, the offset between this magnitude system and the one used by us is negligible.

Our determination of LF agrees well with \citet{AA.342.408G}, within the $68\%$ confidence level (see Fig.~\ref{fig:lf}), and has a value of $M_\star$ that is consistent with that of \citet{AA.367.59P}.
 
The $r-$band total luminosity was calculated to $L_{\rm tot}=L_{\star}\phi_{\star}\Gamma(2+\alpha)$.
The transformation from absolute magnitudes $M_{\star}$ to absolute luminosity $L_{\star}$, in units of solar luminosities, was performed 
using the solar absolute magnitude, obtained using the color-transformation equation from the Johnson-Morgan-Cousins system to the SDSS system of \citet{PASP.107.945F}.
The errors were estimated by the propagation of the $68\%$-confidence-errors of each parameter. 
In this way we found that $L_{\rm tot}=(80\pm2) \times 10^{11}$ $L_\odot$, 
which corresponds to $M_{200}/L_{\rm tot} \sim 230$ ($M_{200} = 1.8 \times 10^{15}$ $M_\odot$, Table~\ref{tab:masses}). 
\citet{AA.464.451P} found a relation between $M_{200}$ and the $r-$band luminosity in 217 clusters selected from 
the  Sloan Digital Sky Survey (see their Eq.~6): according to this relation, the luminosity expected for 
$M_{200} = 1.8 \times 10^{15}$  $M_{\odot}$ was  $L = (73\pm10) \times 10^{11}$ $L_\odot$, 
in excellent agreement with the observed value.

\section{Conclusions}
\label{sec:conclusions}  
For the galaxy cluster Abell 2163, we have shown that by the usage of wide-field imaging it is possible to achieve far better agreement than before, between mass and velocity dispersion measured using weak-lensing and those derived for example from X-ray data.

The dispersion velocity here measured, $\sigma =  1139^{+52}_{-55}$ km s$^{-1}$, agrees well with those derived by X-ray and spectroscopic data, as found by \citet{apj...613...95c}, whereas it was underestimated in the previous analysis by \citet{apj...482..648s}. 

On the other hand, the comparison with the masses obtained from X-ray measurements \citep{apj...456..437m} shows that at $r ~ \sim 2\ {\rm Mpc}$,  
$M_X \sim (1.5 \pm 0.2)   \times 10^{15}$ $M_{\odot}$ ($h=0.7$); at the same distance, we obtain from the NFW fit (Sect.~\ref{subsubsec:models}) 
$M_{\rm wl} \sim (1.6 \pm 0.3)  \times 10^{15}$ $M_{\odot}$. We therefore agree with \citet{apj...482..648s} about the consistency of the mass obtained by 
weak lensing and X-rays: no correction factor is required in our case due to the larger field of view. 
We also find a substantial agreement between our estimate of the virial mass, $M_{\rm vir} = (2.2 \pm 0.4) \times 10^{15}$ $M_\odot$ using the NFW fit, and the value $M_{\rm vir} = (3.8 \pm 0.4) \times 10^{15}$ $M_\odot$ obtained by \citet{maurogordato08} from optical and spectroscopic data; in addition, as noted by these authors, their estimate of the virial mass could be overestimated by 25\%. 
Our weak-lensing analysis also confirms that the mass
distribution is extended along the E-W direction, in agreement with that observed  in optical, radio and X-ray data \citep[see e.g.][]{maurogordato08}.

Finally, the $r-$band total cluster luminosity within $r_{200}$, derived from the luminosity function, gives $L_{\rm tot}/M_{200} = 240$.
The observed luminosity is in very good agreement with that expected for the mass measured by weak lensing, according to the $L-M$ relation proposed by \citet{AA.464.451P}.

\acknowledgement{We warmly thank Thomas Erben for having provided us the software for the KSB analysis. E. Puddu thanks S. Andreon for useful suggestions and comments about the LF determination. We are grateful to the referee for his comments, which improved the paper.
This research is based on observations  made with the Canada-France Hawaii Telescope obtained 
 using the facilities of the Canadian Astronomy Data Centre operated by the National Research Council of Canada with the support of the Canadian Space Agency. This research was partly based on the grant PRIN INAF 2005.
}

\bibliography{9731}

\end{document}